\documentclass{emulateapj}

\usepackage{natbib}
\usepackage[normalem]{ulem}
\shorttitle{3C~279 outburst in 2015: SED study}
\shortauthors{Bottacini et al.}

\begin{document}

\title{\object{3C~279} in Outburst in June 2015: A Broadband-SED Study Based on the {\em INTEGRAL} Detection}
\author{Eugenio~Bottacini\altaffilmark{1}, Markus~B{\"o}ttcher\altaffilmark{2}, Elena~Pian\altaffilmark{3,4,5} and Werner~Collmar\altaffilmark{6}}

\altaffiltext{1}{W.W. Hansen Experimental Physics Laboratory \& Kavli Institute for Particle Astrophysics and Cosmology, Stanford University, USA}
\altaffiltext{2}{Centre for Space Research, North-West University, Potchefstroom 2531, South Africa}
\altaffiltext{3}{INAF, Istituto di Astrofisica Spaziale e Fisica Cosmica, via P. Gobetti 101, 40129 Bologna, Italy}
\altaffiltext{4}{Scuola Normale Superiore, Piazza dei Cavalieri 7, 56122 Pisa Italy}
\altaffiltext{5}{INFN, Sezione di Pisa, Largo Pontecorvo 3, 56127 Pisa, Italy}
\altaffiltext{6}{Max-Planck-Institut f{\"u}r extraterrestrische Physik, Giessenbach, 85748 Garching, Germany}

\begin{abstract}
Blazars radiate from radio through gamma-ray frequencies thereby being ideal targets for multifrequency studies.
Such studies allow constraining the properties of the emitting jet. 3C~279 is among the most notable blazars
and therefore subject to extensive multifrequency campaigns. 
We report the results of a campaign ranging from near-IR to gamma-ray energies of an outburst of 3C~279 in June 2015. The campaign pivots around the detection in only 50 ks by {\em INTEGRAL}, whose IBIS/ISGRI data pin  down the high-energy spectral energy distribution component between {\em Swift}-XRT data and {\em Fermi}-LAT data.
The overall spectral energy distribution from near-IR to gamma rays can be well represented by either a leptonic or a lepto-hadronic radiation transfer model. Even though the data are equally well represented by the two models, their inferred parameters challenge the physical conditions in the jet. In fact, the leptonic model requires parameters with a magnetic field far below equipartition with the relativistic particle energy density.  On the contrary, equipartition may be achieved with the lepto-hadronic model, which however implies an extreme total jet power close to Eddington luminosity.
\end{abstract}

\keywords{accretion, accretion disks --- galaxies: jets --- galaxies: individual (3C 279) --- 
radiation mechanisms: non-thermal --- X-rays: galaxies}

\section{Introduction}
Blazars are radio-loud, jet-dominated active galactic nuclei (AGN) with their jets closely aligned
with the line of sight to the observer. Due to this geometric arrangement, their emission is strongly Doppler
enhanced. Consequently, they are bright sources from radio wavelengths to gamma-ray energies allowing for detailed
multifrequency studies. Precise modeling of the multiwavelength emission allows us to constrain the physical 
properties of the emitting source. However, the emission from blazars is known to be variable on various time scales
\citep{Albert07, Aharonian07}. Therefore, for a 
meaningful interpretation of these properties and emission mechanisms, the observations need to be 
at least quasi-simultaneous and covering a wide range of frequencies. Such a detailed multifrequency campaign
was carried out on the prominent blazar 3C~279. The campaign pivots around its detection by the {\em INTEGRAL} mission
during an outburst in June 2015. This outburst exhibited the source's brightest flare ever
observed at GeV energies \citep{cutini15, lucarelli15, paliya15}. 3C~279, at redshift $z$=0.5362 \citep{marziani96},
is among the most luminous and variable extragalactic objects and gained prominence due to its bright flaring state
detected by the Energetic Gamma Ray Experiment Telescope (EGRET) at the beginning of the
{\em Compton Gamma-Ray Observatory} (CGRO) mission \citep{hartman92}, and it was the first
(of currently only 5) flat-spectrum radio quasar detected by ground-based Atmospheric Cherenkov Telescope
facilities in very-high-energy (VHE: $E > 100$~GeV) gamma rays \citep{Albert08, mirzoyan15}.
An interesting finding regarding
3C~279 is the change of the optical polarization angle following a gamma-ray flare \citep{abdo10} seen also
in a number of other blazars as shown in a recent study by \cite{kiehlmann16}. Therefore, this
intriguing source is subject to extensive monitoring programs revealing low-activity and high-activity states. 
In a low-activity state, the modeling of the  broad band spectral energy distribution (SED) at X-ray energies is quite satisfactory \citep[e.g.][]{collmar10}, while in some cases in a high-activity state of the source the same modeling is 
challenging \citep[e.g.][]{hayashida15}.\\
In this work, we report the detection in the hard X-ray band of an outburst of 3C~279 around which a multifrequency
campaign is centered. First, we detail the multifrequency campaign, followed by the compilation of the broad band
SED and its modeling.

\section{Observations and Data Analyses of 3C~279}
\subsection{X-ray Observations}
\begin{figure}[ht]
\epsscale{1.00}
\includegraphics[width=0.45\textwidth]{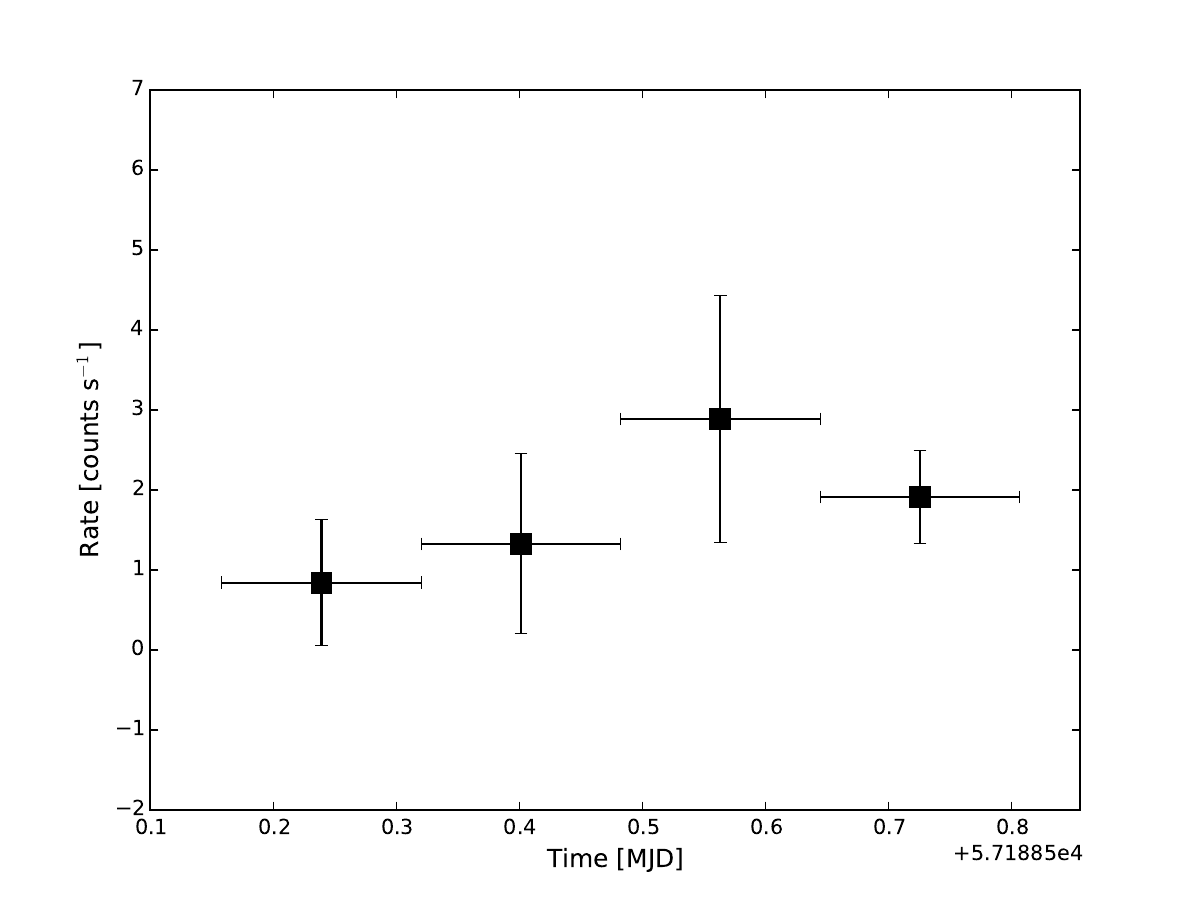}
\caption{IBIS/ISGRI light curve of 3C 279 binned to 12.5 ks.}
\label{fig:ibis-lc}
\end{figure}

\begin{figure}[ht]
\epsscale{1.00}
\includegraphics[width=0.45\textwidth]{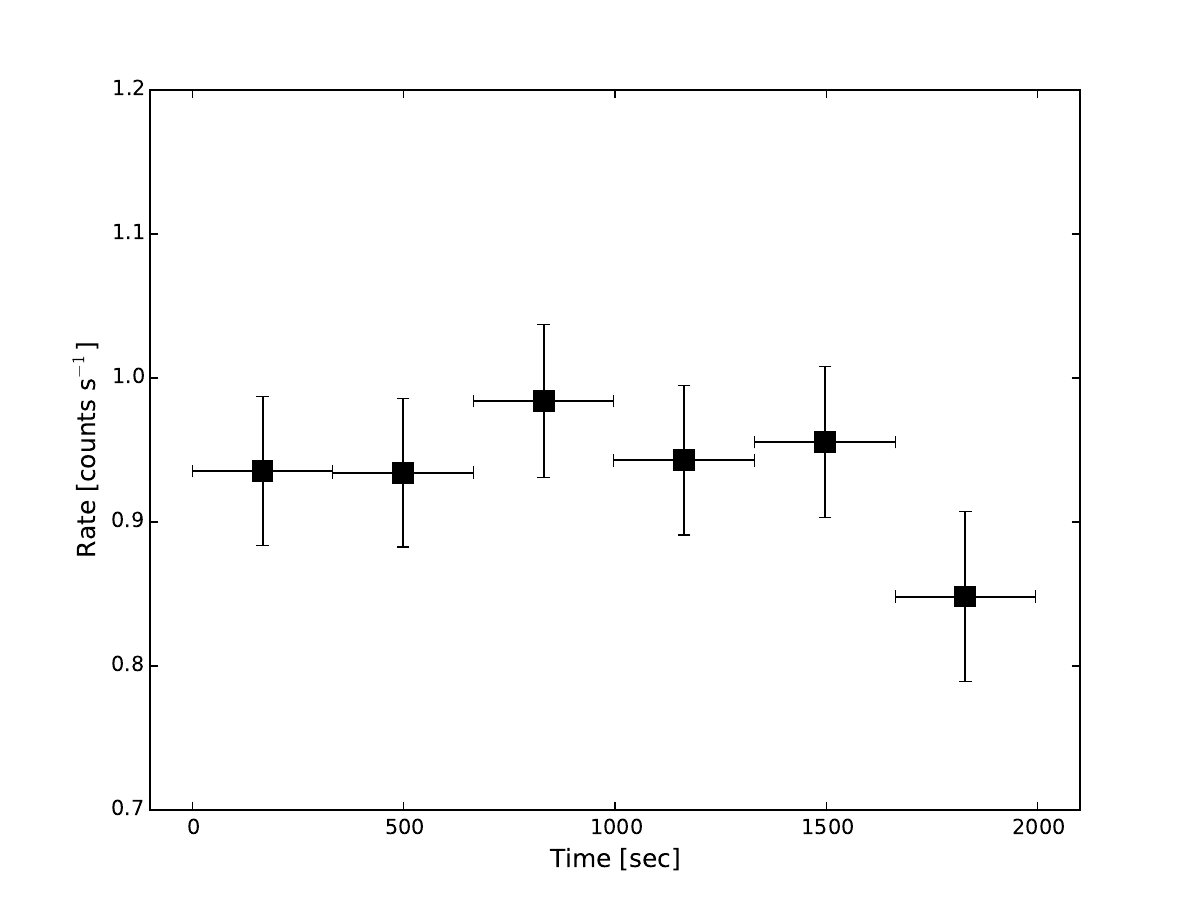}
\caption{{\em Swift}-XRT light curve of 3C 279 binned to $\sim$ 300 s.}
\label{fig:xrt-lc}
\end{figure}

{\em INTEGRAL} -- A monitoring campaign on 3C~279
was performed by the {\em INTEGRAL} mission during its extragalactic survey program on the Coma sky area. While
surveying the sky, astrophysical sources are monitored by {\em INTEGRAL}'s instruments due to their wide field
of view (FOV). This approach of monitoring has provided excellent opportunities for single source studies \citep[e.g.][]{bottacini10}.
During the Coma survey observations in June 2015, {\em INTEGRAL} has detected 3C~279 \citep{bottacini15} 
with its imager IBIS \citep{ubertini03} using the ISGRI detector \citep{lebrun03}. This is IBIS's 
low-energy detector sensitive to photons of 15 -- 1000 keV and having a FOV of 29 $\times$ 29 deg$^2$.
IBIS/ISGRI data of this observation are analyzed with the latest version 10.2 of the standard
Off-line Scientific Analysis (OSA) software\footnote{http://www.isdc.unige.ch/integral/analysis\#Software}
provided by the {\em INTEGRAL} Science Data Centre \citep{courvoisier03} using the method described in \cite{bottacini12}. With this software we analyze
the {\em INTEGRAL} data set of revolution 1553 comprised of 15 Science Windows, each of which corresponds
to a $\sim$3500 seconds pointed observation. These add up to a total exposure time of $\sim$50 ks. Only one 
Science Window (155300040010) has a much shorter exposure of ~500 seconds. The quality of this single
observation compares to all the other Science Windows. The id of the analyzed Science Windows
run from 155300040010 (2015-06-15 15:46 UTC) to 155300170020 (2015-06-16 05:13 UTC) accounting for
pointed observations only. In the resulting mosaic sky image, thanks to its wide FOV, IBIS/ISGRI is able to
detect 3C~279 even at far off-axis angle (14 deg) with respect to the center of the dither pattern at 5.7$\sigma$
in the 18 -- 55 keV energy range at a constant flux level. The source spectrum in the 20 -- 100 keV
energy range can be fit with a flux-pegged power-law model \texttt{pegpwrlw} of XSPEC \citep{arnaud96}, which allows for normalization by total flux, with a 
spectral index of 1.08$^{1.98}_{0.15}$. The result is
reported in Table~\ref{tab:3c279-fit}. 
A timing analysis of this {\em INTEGRAL} observation shows a rather constant
flux displayed in the light curve of Figure~\ref{fig:ibis-lc}. Unfortunately, due to the center of the dither pattern of this 
{\em INTEGRAL} survey on the Coma region being at a large angular distance from 3C~279, the smaller 
FOV of JEM-X and OMC did not allow for an observational coverage of the source at X-ray and optical
frequencies, respectively. Moreover, we have analyzed IBIS/ISGRI data of {\em INTEGRAL}-revolution
1545 including Science Windows from 154500300010 (2015-05-26 04:16 UTC) to 154500470010
(2015-05-26 23:13 UTC), which did not yield a detection of 3C~279. However, the non-detection is used
in the discussion in section 4.
\\
{\em Swift}-XRT -- During the {\em INTEGRAL} monitoring of 3C~279, the source was observed nearly
simultaneously by {\em Swift}-XRT in photon counting mode on 2015-06-15 14:27 UTC (obs id: 00035019176).
The observation log is reported in Table~\ref{tab:3c279-fit}. Data analysis is performed using \texttt{xrtproducts} and
{\em HEAsoft 6.17}. In agreement with the very timely analysis by \cite{pittori15a}, we find the spectrum to be affected by pile-up. To correct for this effect we extract the spectrum from an annulus region (rather than from a 
circular region) that excludes the central circle area of 2 pixels radius. The outer radius of the annulus is 30 pixels 
and the background is extracted from a source-free circular region of 50 pixels \citep[each pixel corresponds to
$\sim$2.36~$\arcsec$,][]{moretti04}. The background spectrum accounts for less than 1\% of the total net count rate.
We generated the ancillary response file (ARF) with \texttt{xrtmkarf} and we used the response matrix of swxpc0to12s6\_20130101v014.rmf \footnote{http://heasarc.gsfc.nasa.gov/docs/heasarc/caldb/data/swift/xrt/index.html}.
To confidently use the chi-square statistic for spectral fitting, we bin the data to at least 20 counts bin$^{-1}$. The
spectrum is fitted in the energy range
0.6 -- 6.0 keV. The baseline fit model is a power law with neutral hydrogen absorption (\texttt{wabs*powerlaw}) fixed
to the Galactic value, which is N$^{gal}_{H}$ = 2.2$\times$10$^{20}$ atoms cm$^{2}$ obtained from the LAB Survey of 
Galactic HI database \citep{kalberla05}. We do not find any evidence for excess absorption or for more sophisticated
models. The fit results and derived values, which are in agreement with the analysis by \cite{pittori15a}, are
reported in Table~\ref{tab:3c279-fit}. No significant variability within this observation is found (see light curve in Figure~\ref{fig:xrt-lc}). Additionally, we have
analyzed {\em Swift}-XRT observation 00092194008 (2015-05-26 09:20 UTC), which is used in the discussion in
section 4.
\\
\subsection{UV to Near-IR Observations}
{\em Swift}-UVOT -- Simultaneously to the {\em INTEGRAL} and {\em Swift}-XRT observations, the UV--Optical Telescope
\citep[UVOT][]{roming05} on board the {\em Swift} satellite observed 3C~279 with its filter of the day, which
was the $U$ filter. 
We use the standard pipeline products, which are co-added and corrected for exposure. The standard 5~$\arcmin$ 
aperture is used for photometry. The resulting magnitude is ${U}$=14.92 ($\pm$ 0.03) that is in agreement with the detailed analysis
in \cite{pittori15b}. To convert magnitudes to fluxes we use the constants of the photometric system in \cite{poole08}.\\

{\em SMARTS} -- The blazar 3C~279 is part of an extensive monitoring program run by Yale University with the
Small and Moderate Aperture Research Telescope System (SMARTS). During the {\em INTEGRAL}
monitoring, SMARTS has been monitoring the source quasi-simultaneously.
Data reduction and analysis are detailed in \cite{bonning12}.
Observations are performed in the optical and near-IR bands $B, V, R, J,$ and $K$ (see Table~\ref{tab:optical}).
Similarly to a precise study by \cite{larinov08}, we convert magnitudes to fluxes using the constants from \cite{mead90}.\\
Correction for the effect of interstellar extinction of {\em Swift}-UVOT and {\em SMARTS} collected data is presented in section 3.

\subsection{Gamma-ray Observations}
Simultaneous to the {\em INTEGRAL} observations, 3C~279 was also observed at MeV and GeV gamma-ray
energies. The detection by the Italian {\em AGILE} mission \citep{tavani08} is detailed in \cite{lucarelli15},
and the {\em Fermi} gamma-ray mission has detected 3C~279 \citep{cutini15} simultaneously to
{\em INTEGRAL} at GeV energies with its Large Area Telescope \citep[LAT;][]{atwood09}.
A prompt analysis of this detection of 3C~279 was performed in a precise time-resolved
spectral study by \cite{paliya15}. From this study, we take the LAT spectral results for the
time window that is contemporaneous to the IBIS/ISGRI observations and that shows resolved variability
on 6 hours time scale.

\section{Spectral Energy Distribution and its Modeling}
Even though the line of sight to the source is hardly affected by extinction due to its position at high Galactic
latitude, we correct for this attenuation in the near-IR, optical and UV bands (i.e. SMARTS and UVOT data). We
compute the extinction with Cardelli's law \citep{cardelli89} with a reddening coefficient of $R_{V}$ = 3.1
and an extinction coefficient in the $V$ band of $A_{V}$ = 0.095 \citep{schlegel98}.\\
The SED of 3C~279 displays the two broad non-thermal radiation components 
characteristic of blazars. The low-energy component is well understood as being due to synchrotron emission
by relativistic electrons (possibly also positrons) in the jet, while the high-energy component can be either due to Compton scattering
by the same relativistic electrons (leptonic processes) or due to proton synchrotron radiation and synchrotron
emission from secondaries produced in photo-pion interactions (hadronic processes). The peak position of the 
high energy component is well constrained by the LAT data: the best-fit \texttt{log-parabola} spectrum
having a spectral index $\alpha$=2.05 ($\pm 0.05$) indicates the peak of the SED to be at $\nu_{\rm HE} \lesssim
3 \times 10^{22}$~Hz \citep{paliya15}. The onset of the high-energy component is constrained by the {\em Swift}-XRT data. 
The onset and the peak of the high-energy component are bridged by the IBIS/ISGRI data.
At low energies, the synchrotron component is constrained by the {\em Swift}-UVOT and SMARTS data,
placing its peak at frequencies below $\nu_{\rm sy} \lesssim 10^{14}$~Hz.\\
To model the overall SED, we adopt the two complementary approaches mentioned above: a leptonic and
a hadronic model. Their time-independent homogeneous 
one-zone jet radiation transfer is used as described in detail in \cite{boettcher13}, 
building upon the earlier work of \cite{BMS97} and \cite{BC02}. In both models, a spherical (in the co-moving frame)
emission region of radius $R$ moves with bulk Lorentz factor $\Gamma$ (corresponding to a speed of 
$\beta_{\Gamma }c$) within an angle $\theta_{\rm obs}$ determining the Doppler factor $\delta = 
\left( \Gamma \, [1 - \beta_{\Gamma} \, \cos\theta_{\rm obs}] \right)^{-1}$. The emitting volume is 
pervaded by a randomly oriented magnetic field of strength $B$ and it is subject to injection of relativistic 
leptons (and protons, in the case of a hadronic model) with a comoving power-law distribution in particle 
energies ($i = e, p$ for electrons/positrons and protons, respectively):
\begin{equation}
\begin{array}{c}
Q_i^{\rm inj} (\gamma; t) = Q_{i,0}^{\rm inj} (t) \, \gamma_i^{-q} \; \; [cm^{-3} s^{-1}] \; \; \\
{\rm for} \;\; \gamma_{i,1} \le \gamma_i \le \gamma_{i,2}
\label{Qe}
\end{array}
\end{equation}
which may be thought of as the result of an unspecified, rapid acceleration mechanism, such as 
first-order Fermi acceleration at a shock traveling through the jet.\\

\begin{figure}[ht]
\epsscale{1.00}
\includegraphics[width=0.5\textwidth]{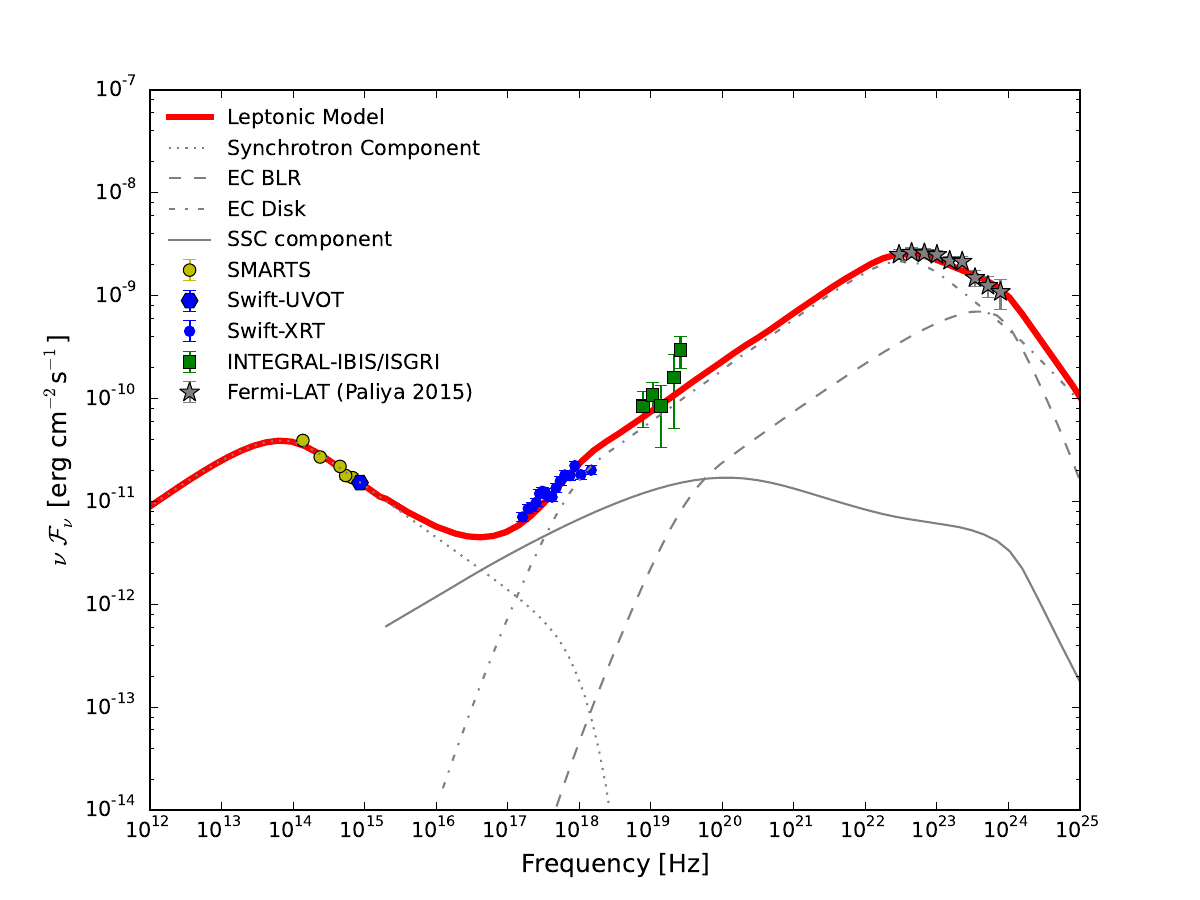}
\caption{Quasi-simultaneous SED of 3C~279, along with the leptonic (red solid) model, as described in the text.
In the leptonic model, the different line styles represent the individual
emission components: synchrotron (dotted), synchrotron self-Compton (solid), external Compton
scattering of accretion disk photons (dot-dashed), external Compton scattering of BLR radiation
(dashed), and the sum of all components (solid). See Table \ref{tab:tab-sed}
for model parameters.}
\label{fig:sed-leptonic}
\end{figure}

\begin{figure}
\epsscale{1.00}
\includegraphics[width=0.5\textwidth]{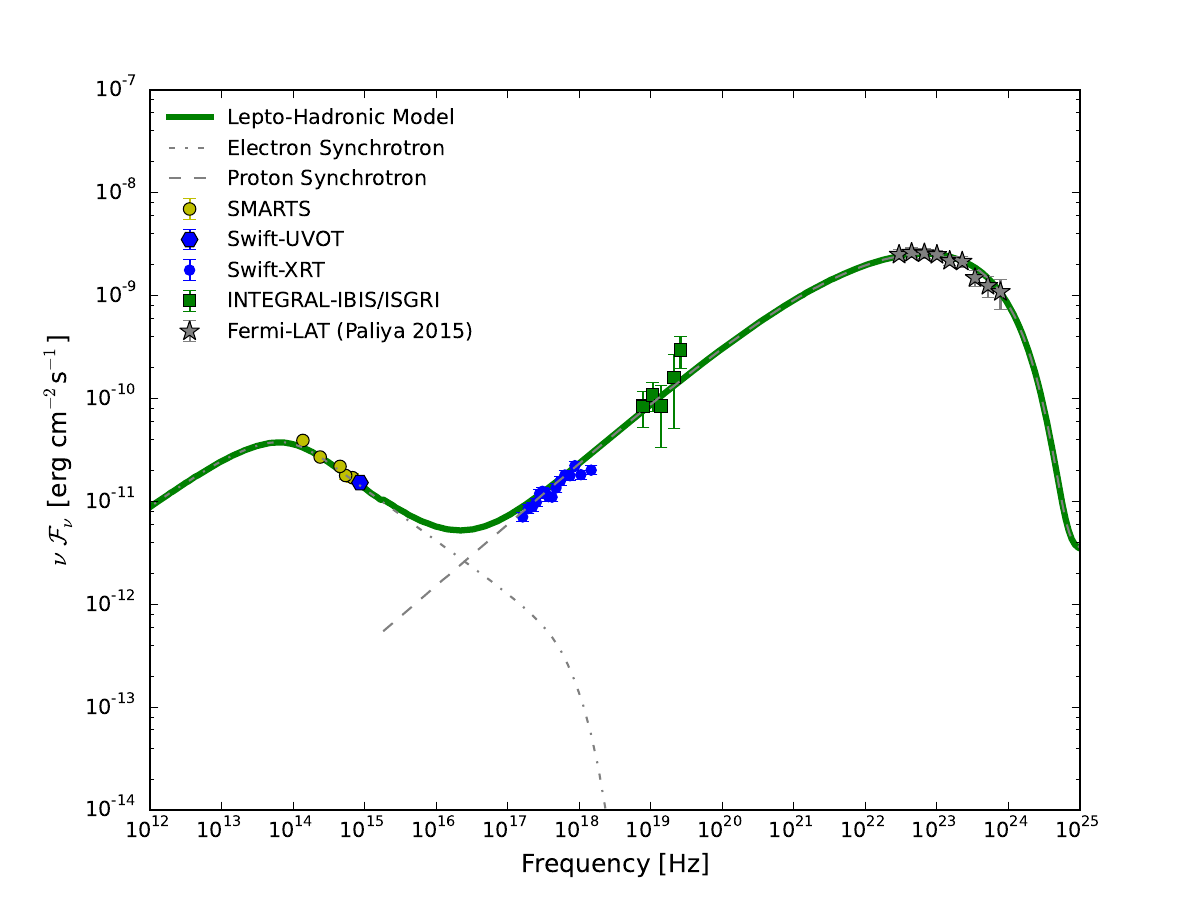}
\caption{Quasi-simultaneous SED of 3C~279, along with the lepto-hadronic (green solid) model,
as described in the text. In the lepto-hadronic model, the different components are:
primary electron synchrotron (dot-dashed) and proton synchrotron (dashed). See Table \ref{tab:tab-sed} for model parameters.}
\label{fig:sed-hadron}
\end{figure}

In the leptonic model, the electron cooling is driven by synchrotron and inverse-Compton emission. 
For the latter the code accounts for synchrotron self-Compton (SSC) and external radiation fields from 
the accretion disk (EC-disk) known to be present in this source \citep[e.g.][]{pian99, hartman01} and an isotropic external radiation field, which represents the emission from
the broad line region (BLR -- EC-BLR) or an infrared-emitting dust torus (EC-DT). A detailed description 
can be found in \cite{boettcher13}. The solution for the equilibrium state between particle injection, cooling, 
and escape from the spherical volume is self-consistently computed. The result is illustrated by
the red solid curve in Figure~\ref{fig:sed-leptonic}, and the model parameters are reported in Table~\ref{tab:tab-sed}. 
In this model, the {\it Swift}-XRT X-ray emission is produced by a combination of SSC and EC-disk emission.
\\
The observational evidence for the existence of this latter component in the X-ray regime of 3C~279 has been set
forth in an early multifrequency study of 3C~279 by \cite{pian99}. Even though in that work the source was in a low-activity
state, the same EC-disk component alone might dominate the hard X-ray emission in the {\em INTEGRAL} regime.
The {\em Fermi}-LAT spectrum is a combination of the EC-disk radiation dominating at lower energies and the
EC-BLR emission dominating above a few GeV. 
The slope of the injected electron distribution of 3 can plausibly be obtained, e.g.,  in diffuse shock acceleration at oblique relativistic shocks \citep[e.g.][]{SB12}. The photon energy densities due to the accretion disk and of the isotropic external radiation field in the stationary rest-frame of the AGN are 4.9 erg cm$^{-3}$ and 3.0$\times$10$^{-3}$ erg cm$^{-3}$, respectively. The inferred location of the emitting region is 0.011 pc from the SMBH.
The resulting power in Poynting flux obtained from a magnetic 
field $B$=1 G is $L_{B}$=2.0$\times$10$^{44}$ erg s$^{-1}$ and the kinetic power carried in relativistic 
electrons is $L_{e}$=2.1$\times$10$^{45}$ erg s$^{-1}$. This corresponds to a magnetization parameter of 
$\epsilon_B \equiv L_{B}/L_{e} = 0.095$, indicating that the emission region is kinetic-energy dominated 
by a factor of $\sim$~$10$. These modeling results are underpinned by the observationally well constraint
accretion disk and BLR parameters.\\
In the lepto-hadronic model, in addition to the leptonic processes described above, the contribution
of ultrarelativistic protons to the radiative output is accounted for. The protons are subject to synchrotron radiation
and photo-pion production. As hadronic models require significantly higher magnetic fields (typically
of order $\sim$~$100$~G) than leptonic ones, electron cooling is strongly dominated by synchrotron emission, and the dominant target photon field for photo-pion production by the relativistic protons is the co-moving synchrotron radiation of the primary electrons with a photon energy density $u'_{syn}$=2.5$\times$10$^{-2}$ erg cm$^{-3}$ compared to the magnetic field energy density $u'_{B}$=570 erg cm$^{-3}$. The assumed dominance of the synchrotron emission over the external radiation photon field, places the emitting region necessarily outside the BLR to satisfy the requirement of $u'_{syn}$ $>$ $\Gamma^{2} \times u_{ext}$.
The equilibrium particle  (electron/positron and proton) spectra and the radiative outputs are calculated self-
consistently, as described in \citep{boettcher13}. The result is shown by the green solid curve in Figure
\ref{fig:sed-hadron}, and the model parameters are listed in Table~\ref{tab:tab-sed}. In the lepto-hadronic
model, the entire X-ray through gamma-ray emission is strongly dominated by proton synchrotron
radiation, with negligible contribution from photo-pion induced processes.
\\
The model requires
a magnetic field of $B = 100$~G, corresponding to a power carried in Poynting flux of $L_B = 1.5
\times 10^{48}$~erg~s$^{-1}$, compared to a power carried in relativistic protons of $L_p = 1.6
\times 10^{48}$~erg~s$^{-1}$ and in relativistic leptons of $L_e = 1.9 \times 10^{42}$~erg~s$^{-1}$.
It is obvious that the kinetic energy in the jet is strongly dominated by the relativistic
protons, and the magnetization parameter, i.e., the ratio of magnetic-field to kinetic energy
carried in the jet, is $\epsilon_B \approx L_B / L_p = 0.9$. This indicates that in this model,
kinetic and magnetic energy are significantly closer to equipartition than in the case of our
leptonic model.
Additionally, for the external radiation field to be negligible, it requires the emission region to be located outside the BLR, which will also allow for potential VHE gamma-ray emission to escaped unattenuated.
Figure~\ref{fig:sed-leptonic} and Figure~\ref{fig:sed-hadron} illustrate that both (leptonic and lepto-hadronic) models are able to almost indistinguishably represent the SED data, and both models are able to reproduce the spectral curvature
seen in the {\em Fermi}-LAT gamma-ray spectrum well. However, the lepto-hadronic model requires a
very hard spectral index of the proton spectrum of 1.8, which may be difficult to realize in nature.
It is furthermore subject to the well-known extreme power requirements of lepto-hadronic models of
blazars \citep[e.g.][]{boettcher13,DB15,Petropoulou15}, which may be incompatible with our current 
understanding of jet launching mechanisms \citep{Zdziarski15}. Specifically, our model requires a total jet 
power of $L_{\rm jet} \approx 3.1 \times 10^{48}$~erg~s$^{-1}$, which is almost exactly equal to 
the Eddington luminosity of the $2.5 \times 10^8 \, M_{\odot}$ black hole powering the AGN in 3C~279.
This finding regarding the total jet power is strengthened by the equipartition condition known to be
close to the minimal energy requirement.

\section{Discussion}
The active state of 3C~279 in June 2015 allowed for a 5.7$\sigma$ detection with IBIS/ISGRI in 
only $\sim$50 ks of observations. This can be compared to a quiescence state, in which IBIS/ISGRI needs a $\sim$10 times longer exposure ($\sim$500 ks) for a 7.5$\sigma$ detection as show in a multifrequency study by \cite{collmar10}. To explore 
possible hard X-ray variability on longer time scales, we analyzed the data of {\em INTEGRAL} 
revolution 1545 a month earlier, in May 2015, for a total exposure of $\sim$44 ks. This did not
yield a detection, which indicates that the source was in a quiescence state. A confirmation of 
this finding is provided by a simultaneous {\em Swift}-XRT observation (id 00092194008, see 
Table~\ref{tab:3c279-fit}) that reveals the source flux to be three times lower with respect 
to the active state. This draws a consistent picture of the source state and indicates that
the hard X-ray emission of 3C~279 is variable and that this variability may be correlated 
with the soft X-ray variability, as expected in both model scenarios considered here, in 
which the emission in both frequency regimes is produced by the same relativistic particle 
population. 
We find that homogeneous, single-zone radiation transfer models provide a satisfactory representation
of the SED, both in a leptonic and a lepto-hadronic scenario.\\
Our leptonic SED model indicates that the
emission region needs to be kinetic-energy dominated by a factor of $\sim$~$10$ with respect to the 
energy carried in magnetic fields. Such a situation may be difficult to realize in jet formation and
acceleration scenarios in which the magnetic field is the primary source of jet power, as one would
naturally expect any mechanism that converts magnetic-field to particle kinetic energy to cease once
approximate equipartition is reached. A particle acceleration scenario involving relativistic shocks
in a particle-dominated jet \citep[e.g.,][]{Marscher85, Sokolov04, Joshi11} and/or shear layer 
acceleration in a spine-sheath geometry \citep[e.g.][]{SO02, Ghisellini05} may provide a suitable 
alternative.
As for the leptonic scenario the inferred location of the emitting region is well within the BLR (at 0.011 pc from the SMBH), VHE photons would not be able to escape this region unattenuated \citep{boettcher16}. Thus, our model predicts that no VHE emission should have been detected from this flare, if not originating at a different location. Such a location is found to be far beyond the BLR of 3C~279 in order for VHE photons of $\sim$100 GeV to be detectable, as derived in a dedicated study for the gamma-ray emission site by \cite{dermer14}. In fact, the single highest-energy gamma ray revealed by the {\em Fermi}-LAT in this flare from 3C~279 is detected at no more than $\sim$52 GeV \citep{paliya15}.\\
In contrast, our lepto-hadronic SED model allows us to choose parameters close to
equipartition between the magnetic field and the relativistic proton population. This model 
requires a proton injection spectral index of 1.8, which is harder than what one would expect,
e.g., in standard scenarios of particle acceleration at non-relativistic shocks or relativistic,
parallel shocks. However, diffusive shock acceleration at relativistic, oblique shocks \citep{SB12}
or at relativistic shear layers \citep[e.g.][]{RD06} may produce relativistic particle spectra
much harder than $E^{-2}$. 
Further support for the lepto-hadronic scenario is provided by the quite good agreement in terms of the
minimum-time variability of $\sim$6.2 h derived from our modeling and the measured resolved variability
of $\sim$6 h in the study of \cite{paliya15}. In our modeling, we have assumed that the co-moving electron-synchrotron radiation field dominates over external radiation fields, which requires the emission region to be located outside the BLR.
An important issue for the lepto-hadronic model may be the extreme jet 
power, of the order of the Eddington luminosity of the central black hole in 3C~279.
Our observations show that {\em INTEGRAL} data tie in with both, the {\em Swift}-XRT soft X-ray data and the {\em Fermi}-LAT
gamma-ray data, forming a straight power law, which strongly suggests that it is one single emission
component that is responsible for the entire X-ray through gamma-ray spectrum. This is compatible with
the proton-synchrotron interpretation.\\
The low-energy synchrotron emission and the high-energy inverse-Compton emission, originate
in the very same region in the assumption of the one-zone radiation transfer. In this assumption,
the leptonic model cannot
explain the observed low-energy and high-energy variability behavior in many cases
\citep[e.g.][]{boettcher09, nalewajko12, paliya16}, which calls for alternative sophisticated models
including the lepto-hadronic model that faces however the issue of the extreme jet power in our study.
Dedicated studies of hadronic processes are also put forward in a sophisticated work by \cite{ackermann16} to explain
sub-orbital scale variability in 2 out of 11 one-orbit bins during a {\em Fermi}-LAT ToO observation.
A time-dependent lepto-hadronic model \citep[e.g.][]{DB15} might represent a viable avenue for a solution.\\
In quiescence states of 3C~279, leptonic models can often reproduce the SED satisfactory \citep[e.g.][]{collmar10, zheng16}. Such modeling suggests that the gamma-ray emission site is outside the BLR \citep{zheng16} and thus making the detection of VHE photons possible. On the other hand the leptonic model is far from equipartition \citep{boettcher09} when trying to reproduce the Major Atmospheric Gamma-Ray Imaging Cherenkov Telescope (MAGIC) VHE data \citep{Albert08}. However, those data can be well explained in the context of a lepto-hadronic model \citep{boettcher09}.

\section{Conclusions}
3C~279 was caught in outburst by {\em INTEGRAL}--IBIS/ISGRI in June 2015. The multifrequency
campaign around this detection has collected observations in the optical, the X-ray, the hard
X-ray, and the gamma-ray bands. The SED of these observations can be equally well modeled
by a leptonic and by a lepto-hadronic model. Yet, the derived parameters of these models challenge  
the physical conditions in the jet. In fact, for the leptonic model an even approximate to equipartition
arrangement cannot be obtained. The same model also predicts that no VHE photons should have been detected during this outburst. On the contrary a close to equipartition arrangement in the lepto-hadronic
scenario favors this model, even though the jet power remains extreme. Our {\em INTEGRAL}
data tie in {\em Swift}-XRT and {\em Fermi}-LAT data forming a straight power law, which is compatible with the proton-synchrotron component of the lepto-hadronic scenario.

\acknowledgments
We thank the anonymous referee for an exhaustive and constructive refereeing process.
We also thank the {\em INTEGRAL} and the {\em Swift} teams for the observations.
This research is based on observations with {\em INTEGRAL}, an ESA project with instruments and science
data centre funded by ESA member states (especially the PI countries: Denmark, France,
Germany, Italy, Switzerland, Spain), and Poland, and with the participation of Russia and
the USA.
This paper has made use of up-to-date SMARTS optical/near-infrared light curves that are available at
www.astro.yale.edu/smarts/glast/home.php. 
This research has made use of the NASA/IPAC extragalactic Database (NED), which
is operated by the Jet Propulsion Laboratory, of data obtained from the High Energy
Astrophysics Science Archive Research Center (HEASARC) provided by NASA's
Goddard Space Flight Center.
E.P. acknowledges partial support from the Italian Ministry for Research and
Scuola Normale Superiore, and grants ASI INAF I/088/06/0, INAF PRIN 2011, PRIN MIUR 2010/2011.
The work of M.B. is supported by the South African Research Chair Initiative (SARChI) of the Department
of Science and Technology and the National Research Foundation\footnote{Any opinion, finding, and conclusion
or recommendation expressed in this material is that of the authors and the NRF does not accept any liability
in this regard.} of South Africa through SARChI Chair grant No. 64789.
E.B. acknowledges support through NASA grants NNX13AO84G and NNX13AF13G.\\
In memory of Lea Bottacini.

{\it Facilities:} \facility{Fermi}, \facility{INTEGRAL}, \facility{Swift}, \facility{SMARTS}

\clearpage
\begin{deluxetable}{ccccccccc}
\tablewidth{0pt}
\tabletypesize{\scriptsize}
\tablecaption{3C~279: Hard X-ray and X-ray Spectral Fits. \label{tab:3c279-fit}}
\tablehead
{
\colhead{Observation} 			&\colhead{Start}				&
\colhead{Expo} 					&\colhead{Model}				&
\colhead{$\Gamma$}    			&\colhead{Norm} 				&
\colhead{Chi-square} 			&\colhead{d.o.f.} 				&
 \colhead{Flux} \\
\colhead{\scriptsize [$Inst. ~obs.id$]} 			& \colhead{\scriptsize [$date~UTC$]}		&
\colhead{\scriptsize [$sec$]}					& \colhead{\scriptsize}					&
\colhead{\scriptsize}		 					& \colhead{\scriptsize} 					& 
\colhead{\scriptsize} 							& \colhead{\scriptsize}					&	
\colhead{\scriptsize[$10^{-11}~erg~cm^{-2}~s^{-1}$]}
}
\startdata
IBIS 12200240002 & 2015-06-15 15:46:00 & 50000 & pegpwrlw & 1.08$^{1.98}_{0.15}$ & 1.62$^{2.16}_{1.09}$ & 4.79 & 5 & 7.83$^{10.45}_{5.57}$ \\
XRT 00035019176 & 2015-06-15 14:27:58 & 1995 & abs*powerlaw & 1.37$^{1.44}_{1.30}$ & 5.84$^{6.20}_{5.48}$ & 32.22 & 35 & 2.27$^{2.37}_{2.19}$ \\
XRT 00092194008 & 2015-05-26 09:20:59 & 1115 & abs*powerlaw & 1.85$^{1.99}_{1.71}$ & 3.35$^{3.69}_{2.99}$ & 14.33 & 19 & 0.703$^{0.705}_{0.640}$ \\
 \enddata
\tablecomments{Normalization: IBIS/ISGRI in units of 10$^{10}~erg~cm^{-2}~s^{-1}$ in the 20--100 keV band. {\em Swift}-XRT in units of $10^{-3}~ph~keV^{-1}~cm^{-2}~s^{-1}$. Fluxes: IBIS/ISGRI flux is given in the 18--55 keV band. {\em Swift}-XRT flux is computed in the 2--6 keV band. Errors are reported at 1$\sigma$ level.}
\end{deluxetable}

\clearpage
\begin{deluxetable}{ccccc}
\tablewidth{0pt}
\tabletypesize{\scriptsize}
\tablecaption{UV to Near-IR Observations. \label{tab:optical}}
\tablehead
{
\colhead{Instrument}		&
\colhead{Start} 			&\colhead{Filter}				&
\colhead{Mag} 			&
\colhead{Mag Error}  \\
\colhead{\scriptsize} 			&
\colhead{\scriptsize [UTC]} 	& \colhead{\scriptsize}			&
\colhead{\scriptsize}			&
\colhead{\scriptsize}
}
\startdata
SMARTS  & 2015-06-15 23:15:24 & B & 15.706 & 0.004\\
SMARTS  & 2015-06-15 23:19:21 & V & 15.187 & 0.004\\
SMARTS  & 2015-06-15 23:22:24 & R & 14.662 & 0.004\\
SMARTS  & 2015-06-15 23:15:20 & J & 12.903 & 0.005\\
SMARTS  & 2015-06-15 23:19:18 & K & 10.651 & 0.003\\
UVOT	& 2015-06-15 14:32:11 & U & 14.92 & 0.03\\

\enddata
\end{deluxetable}


\clearpage
\begin{deluxetable}{lcccccccccc}
\tablewidth{0pt}
\tabletypesize{\scriptsize}
\tablecaption{List of Parameters from SED Modeling. \label{tab:tab-sed}}
\tablehead
{
\colhead{Model} 			& \colhead{$R$}						&	
\colhead{$\eta_{\rm esc}$} 	&
\colhead{$B$}    			& \colhead{$\Gamma$}     				&
\colhead{$\gamma_{\rm min}$} 	& \colhead{$\gamma_{\rm max}$} 			&
\colhead{$q$} 				&\colhead{$\gamma_{\rm pmin}$} 			&
\colhead{$\gamma_{\rm pmax}$}	& \colhead{$q_{\rm p}$} 				\\
\colhead{\scriptsize(1)} 		&\colhead{\scriptsize(2)}		&
\colhead{\scriptsize(3)}		&
\colhead{\scriptsize(4)}		&\colhead{\scriptsize(5)} 		&
\colhead{\scriptsize(6)}              &\colhead{\scriptsize(7)}		& 
\colhead{\scriptsize(8)} 		&\colhead{\scriptsize(9)}		&
\colhead{\scriptsize(10)}            &\colhead{\scriptsize(11)}	
}
\startdata
Leptonic 		& 1 & 7 &  1 & 20 & 1$\times$10$^{3}$ & 1$\times$10$^{5}$ & 3 & \nodata & \nodata & \nodata\\
Lepto-hadronic & 1 & 1 &  100 & 20 & 1$\times$10$^{2}$ & 1$\times$10$^{4}$ & 3 & 1$\times$10$^{3}$ & 6$\times$10$^{8}$ & 1.8\\
\enddata
\tablecomments{Explanation of columns: (1) SED model; (2) radius of emitting volume $\times$ 10$^{16}$ cm; (3) escape time parameter: t$_{\rm esc}$=$\eta_{\rm esc}$$\times R/c$; (4) magnetic field in Gauss; (5) bulk Lorentz factor; (6) and (7) injected electrons minimum and maximum random Lorentz factors; (8) slope of injected electron distribution; (9) and (10) low and high energy cut-off of proton spectrum in GeV; (11) slope of injected proton distribution.}
\end{deluxetable}


\end{document}